# Highly nonlinear magnetoelectric effect in antiferromagnetic $Co_4Ta_2O_9$ single crystals

Nara Lee,[1] Dong Gun Oh,[1] Sungkyun Choi,[2,3] Jae Young Moon,[1] Jong Hyuk Kim,[1] Hyun Jun Shin,[1] Hwan Young Choi,[1] Kwanghyo Son,[4] Matthias J. Gutmann,[5] Gideok Kim,[2] Jürgen Nuss,[2] Valery Kiryukhin[3] and Young Jai Choi[1,*]

[1]*Department of Physics, Yonsei University, Seoul 03722, Korea*
[2]*Max Planck Institute for Solid State Research, Heisenbergstrasse 1, 70569 Stuttgart, Germany*
[3]*Department of Physics and Astronomy, Rutgers University, Piscataway, New Jersey 08854, USA*
[4]*Department of Modern Magnetic Systems, Max Planck Institute for Intelligent Systems, Heisenbergstrasse 3, D-70569 Stuttgart, Germany*
[5]*Rutherford Appleton Laboratory, ISIS Facility, Chilton Didcot, Oxfordshire OX11 0QX, United Kingdom*

Strongly correlated materials with multiple order parameters provide unique insights into the fundamental interactions in condensed matter systems and present opportunities for innovative technological applications. A class of antiferromagnetic honeycomb lattices compounds, $A_4B_2O_9$ (A = Co, Fe, Mn; B = Nb, Ta), have been explored owing to the occurrence of linear magnetoelectricity. We observe a highly nonlinear magnetoelectric effect on single crystals of $Co_4Ta_2O_9$ (CTO), distinctive from the linear behavior in the isostructural $Co_4Nb_2O_9$. Ferroelectricity emerges primarily along the [110] direction under magnetic fields, with the onset of antiferromagnetic order at $T_N$ = 20.5 K. For in-plane magnetic field, a spin-flop occurs at $H_C \approx 0.3$ T, above which the ferroelectric polarization gradually becomes negative and reaches a broad minimum. Upon increasing magnetic field further, the polarization crosses zero and increases continuously to ~60 $\mu C/m^2$ at 9 T. In contrast, the polarization for a magnetic field perpendicular to the hexagonal plane increases monotonously and reaches ~80 $\mu C/m^2$ at 9 T. This observation of a strongly nonlinear magnetoelectricity suggests that two types of inequivalent $Co^{2+}$ sublattices generate magnetic field-dependent ferroelectric polarization with opposite signs. These results motivate fundamental and applied research on the intriguing magnetoelectric characteristics of these honeycomb lattice materials.

These authors contributed equally to the work: Nara Lee, Dong Gun Oh

\* Correspondence and request for materials should be addressed to Y. J. C. (phylove@yonsei.ac.kr)

**Introduction**

The emergence of novel cross-coupling effects generated by multiple order parameters in a wide range of materials has provided great insights into the interactions that occur in condensed matter systems[1,2]. Prominent examples are magnetoelectric and multiferroic materials where the cross-coupling between electric and magnetic properties has driven intense research to explore fundamental mechanisms responsible for the intrinsic magnetoelectric effects[3-9]. The primary focus of research activity in this field has on the emergence of ferroelectricity from different types of exotic magnetic orders and its dependence on applied magnetic fields. Some studies have also the potential of these materials in applications such as magnetoelectric memory and sensors by engineering their cross-coupling effects[10-13]. Despite the fact that quite a few magnetoelectric or multiferroic materials are known to us, most of them do not feature substantial coupling between structural distortions and magnetic order. Hence, new materials with stronger magnetoelectric coupling are needed for enhancing the feasibility of utilizing their functionalities in device applications.

Materials composed of two dimensional honeycomb lattices have been investigated due to possible occurrence of intriguing physical phenomena such as quantum spin liquid state[14-16] and electronic state with Dirac-like linear dispersion[17-19], and so on. The hexagonal antiferromagnet of $Co_4Nb_2O_9$, which crystallizes in a trigonal $P\bar{3}c1$ structure with two different types of honeycomb layers stacked alternately along the $c$ axis, has recently been in focus for its linear magnetoelectric behavior[6,7,20-22]. In the single crystalline $Co_4Nb_2O_9$, grown by a floating zone method[23], antiferromagnetic order sets in below $T_N \approx 27$ K where the $Co^{2+}$ moments are aligned approximately along the [$\bar{1}$10] direction. Below $T_N$, a linear magnetoelectric effect is observed[24-26]. In this state, an electric polarization appears upon application of an external magnetic field and its magnitude increases linearly with increasing field strength. The measurement of magnetoelectric properties along different crystallographic axes reveals the elements of the magnetoelectric tensor, which is explained by a magnetic structure with lowered symmetry $C2/c'$[21]. Particularly, the presence of off-diagonal elements in the magnetoelectric tensor suggests the formation of toroidal moments[27-29].

Further studies of the magnetoelectric effect in honeycomb lattices were done on the isostructural compound $Co_4Ta_2O_9$ (CTO)[6,30,31]. In CTO, the antiferromagnetic order emerges at $T_N \approx 20$ K, simultaneously with the appearance of a dielectric anomaly and a ferroelectric polarization in applied magnetic fields. It had been believed that below $T_N$, the ferroelectric polarization in CTO increases monotonously under increasing applied magnetic fields, similar to that in $Co_4Nb_2O_9$[20-22]. However, these studies were performed only on polycrystalline samples, in which the physical properties are averaged out over all spatial directions due to a large number of grains of varying orientations. To overcome this challenge, we grew single crystals of CTO by utilizing the conventional flux method[32] to investigate the magnetic and magnetoelectric properties associated with collinear antiferromagnetic order on honeycomb lattices. The single crystalline CTO, in contrast to polycrystalline samples, reveals strongly nonlinear magnetoelectric effect that suggests the existence of two different polarization components originating from inequivalent $Co^{2+}$ sublattices. This suggests that CTO possesses magnetoelectric characteristics which are distinct among $A_4B_2O_9$ (A = Co, Fe, Mn and B = Nb, Ta) compounds[20,30,33-36].

**Results and Discussion**

CTO crystallizes in a centrosymmetric trigonal $P\bar{3}c1$ structure with unit cell dimensions of $a$ = 0.517 nm, and $c$ = 1.413 nm, obtained from the single crystal X-ray diffraction experiment (see Supplementary Information for details). The crystallographic structures viewed from the top and side are depicted in Figs. 1(a) and (b), respectively. Two dissimilar types of honeycomb layers are stacked alternatingly along the $c$ axis. One layer consists of six edge-shared $CoO_6$ octahedra in the same plane, while the other consists of corner-shared octahedra buckled in a zig-zag arrangement around the ring[20]. To examine the magnetic properties of CTO, the temperature ($T$) dependence of the magnetic susceptibility, $\chi = M/H$ ($M$: magnetization, $H$: magnetic field), was measured at $H = 0.1$ T upon warming after zero-field-cooling. The anisotropic $\chi$, obtained for the $H$ along the three distinguishable axes $[\bar{1}10]$, $[110]$, and $[001]$, are shown in Fig. 1(c). For the two in-plane orientations $[\bar{1}10]$ and $[110]$, the $\chi$ exhibits a sharp anomaly at $T_N \approx 20.5$ K, indicating the emergence of antiferromagnetic order. The $T$ dependence of $C/T$ ($C$: specific heat) measured at zero $H$ also shows a distinct anomaly at $T_N$ (Fig. 1(d)). Above $T_N$, the $\chi$ for the two in-plane orientations decrease smoothly with $T$ with nearly identical shapes. On the other hand, a weak anomaly at $T_N$ is observed in the $\chi$ for the $[001]$ axis. The overall $T$ dependence of $\chi$, compared between in-plane and out-of-plane orientations, shows strong magnetic anisotropy, suggesting the in-plane antiferromagnetic arrangement of $Co^{2+}$ spins. The shape of $\chi$ curve for $[\bar{1}10]$ and $[110]$ axes are different below $T_N$ and the faster decrease of $\chi$ for the $[110]$ axis upon lowering $T$ implies that the spins tend to align more easily along this axis. As $T$ is further decreased, a sudden increase of $\chi$ occurs at $T_C = 6.5$ K. The characteristics of this transition were investigated in detail by AC $\chi$ measurements, which indicates the formation of a short-range ferromagnetic phase accompanied by the possible generation of a glassy state (see Supplementary Information for details).

The isothermal magnetization, $M$, for the three inequivalent orientations was measured up to $\pm 9$ T at $T = 2$ K, as shown in Fig. 2(a). The $M$ along the $[\bar{1}10]$ direction ($M_{[\bar{1}10]}$) shows a broad bending at low $H$ regime, which is compatible with the observation of the short-range ferromagnetic component. Upon increasing $H$ further, the $M_{[\bar{1}10]}$ increases monotonously and reaches 3.9 $\mu_B$/f.u. at 9 T. The $M_{[110]}$ exhibits a similar $H$ dependence to the $M_{[\bar{1}10]}$; however, the magnetic moment at 9 T is found to be ~3.7 $\mu_B$/f.u., which is slightly smaller than that of the $M_{[\bar{1}10]}$. For an applied field along both $[\bar{1}10]$ and $[110]$, the spin-flop transition occurs at $H_C \approx 0.3$ T, which manifests as the change in slope shown in the magnified plot of the $M_{[\bar{1}10]}$ (Fig. 2(b)). Note that this result is different from the previous results on polycrystalline samples where the spin-flop transition occurs at a higher $H$ of ~0.9 T, indicating the averaging effect over grain orientations[31]. The $M_{[001]}$ increases almost linearly up to 9 T resulting in a magnetic moment of ~2.1 $\mu_B$/f.u. at 9 T, consistent with the strong magnetic anisotropy observed in the $T$ dependence of anisotropic $\chi$ (Fig. 1(c)).

The anisotropic characteristics of magnetoelectric properties were examined through the $T$ dependence of $P$ for the [$\bar{1}$10], [110] and [001] axes. The magnitude of $P$ for each orientation was obtained by integrating the pyroelectric current density measured after poling in an electric field along the direction of $P$ and $H$ up to 9 T for the three different orientations, as shown in Fig. 3. The $P$ emerges below $T_N$ and the elements of the magnetoelectric tensor do not vanish similar to the $T$ dependence of $P$ in Co$_4$Nb$_2$O$_9$[21]. However, the magnitude of $P$ along the [$\bar{1}$10] ($P_{[\bar{1}10]}$, Figs. 3(a), (d), and (g)) and [001] ($P_{[001]}$, Figs. 3(c), (f), and (i)) axes are found to be small at 2 K and 9 T, ranging from 1.9 to 18.0 µC/m². A much larger $P$ and intriguing $T$ dependence with increasing $H$ is observed along the [110] axis ($P_{[110]}$). Figure 3(b) shows the $T$-dependence of $P_{[110]}$ at $H$ = 1, 3, 5, 7, and 9 T along the [$\bar{1}$10] axis ($H_{[\bar{1}10]}$). The $P_{[110]}$ at $H_{[\bar{1}10]}$ = 1 T starts from a negative value of −13.2 µC/m² at 2 K, increases monotonously to zero upon increasing $T$, and disappears at $T_N$. At $H_{[\bar{1}10]}$ = 3 T, $P_{[110]}$ exhibits the largest negative value of −32.2 µC/m² at 2 K and crosses zero $P_{[110]}$ at approximately 15 K. A similar trend of change in the sign of $P_{[110]}$ is observed at $H_{[\bar{1}10]}$ = 5 T with an upward shift in the overall magnitude of $P_{[110]}$. The $P_{[110]}$ at $H_{[\bar{1}10]}$ = 7 and 9 T retains positive values throughout the $T$ range below $T_N$ and the maximum observed magnitude of $P_{[110]}$ is 55.9 µC/m² at 2 K and $H_{[\bar{1}10]}$ = 9 T. The strongly nonlinear magnetoelectric behavior is also observed in $P_{[110]}$ at different values of $H_{[110]}$ (Fig. 3(e)). At $H_{[110]}$ = 1 T, the $P_{[110]}$ is very small in magnitude and shows negligible $T$ dependence. The value of $P_{[110]}$ at $H_{[110]}$ = 3, 5, and 7 T are all negative in certain low $T$ regions. In contrast to the case of an in-plane $H$, the $P_{[110]}$ under an applied $H_{[001]}$ tends to increase gradually as $H_{[001]}$ is increased, maintaining a positive value throughout the range of $T$ below $T_N$ (Fig. 3(h)). The $P_{[110]}$ at $H_{[001]}$ = 9 T and 2 K is found to be 78.7 µC/m², which is approximately twice that of $P_{[110]}$ = 34.9 µC/m² at $H_{[110]}$ = 9 T and 2 K (Fig. 3(e)).

Figure 4(a) shows the $T$-dependence of dielectric constant for $E$//[110] ($\varepsilon'_{[110]}$), measured at $H_{[\bar{1}10]}$ = 9 T and $f$ = 100 kHz. The $\varepsilon'_{[110]}$ at 9 T exhibits a very sharp peak at 20.02 K with a 2.8% change in its magnitude (at the peak maximum). The sharpness of the peak at 9 T, characterized by the full width at half maximum (FWHM), is found to be only 0.08 K, which is smaller compared to the FWHM (~0.5 K) obtained from polycrystalline samples[30,31]. As $H$ is decreased, the peak of $\varepsilon'_{[110]}$ shifts progressively to higher $T$ with a gradual reduction of the peak height (Fig. 4(b)) and almost disappears at 4 T. At 5 T, a tiny peak in $\varepsilon'_{[110]}$, with only

0.27 % change in the overall magnitude, occurs at 20.37 K.

The nonlinear behavior of $P$ and the intricate relationship between magnetic and ferroelectric properties were examined in detail by comparing the $H$ dependence of $P$, $M$, and $\varepsilon'$ at 2 K. The isothermal $P_{[110]}$ was obtained by integrating the magnetoelectric current density, measured by sweeping the $H_{[\bar{1}10]}$ between 9 and −9 T at 2 K after poling in $H_{[\bar{1}10]} = 9$ T and $E_{[110]} = 4.72$ kV/cm, as shown in Fig. 5(a). Starting from the maximum value of $P_{[110]} = 52.5$ μC/m² at 9 T, the $P_{[110]}$ decreases upon decreasing $H_{[\bar{1}10]}$ and becomes zero at 6.3 T. As $H_{[\bar{1}10]}$ is decreased further, the $P_{[110]}$ shows a broad minimum at 3.2 T with $P_{[110]} = -27.5$ μC/m². Below $H_C \approx 0.3$ T, $P_{[110]}$ disappears. Further decrease in $H$ in the negative direction leads to the antisymmetric $H$ dependence of the $P_{[110]}$. In other words, $P_{[110]}$ increases towards more positive values above $-H_C$ with a broad maximum at $H_{[\bar{1}10]} = -3.2$ T. The sweeping of $H_{[\bar{1}10]}$ from −9 T to +9 T completes the isothermal $P_{[110]}$ curve, showing negligible magnetic hysteresis. In Fig. 5(b), the magnetodielectric (MD) effect, described by the variation of $\varepsilon'_{[110]}$ by applying $H_{[\bar{1}10]}$ and defined as $\text{MD}_{[110]}(\%) = \frac{\varepsilon'(H) - \varepsilon'(0 \text{ T})}{\varepsilon'(0 \text{ T})} \times 100$, was measured up to ±9 T at $f = 100$ kHz and $T = 2$ K. The initial curve of $\text{MD}_{[110]}$ exhibits a slight curvature at low $H_{[\bar{1}10]}$ regime and the maximum slope at $H_C \approx 0.3$ T. Above $H_C$, $\text{MD}_{[110]}$ is reduces more gradually which becomes almost linear above $H_{[\bar{1}10]} = 1.5$ T. The maximum variation of $\text{MD}_{[110]}$ is found to be approximately −0.36 % at 9 T. The full $\text{MD}_{[110]}$ curve appears to be symmetric because the direction of $P_{[110]}$ is indistinguishable in the AC excitation of $E_{[110]}$ for the $\varepsilon'$ measurement. For a precise comparison with the $\text{MD}_{[110]}$, the $H_{[\bar{1}10]}$ derivative of isothermal $M_{[\bar{1}10]}$, $dM_{[\bar{1}10]}/dH_{[\bar{1}10]}$ at 2 K is also plotted in Fig. 5(c). The $dM_{[\bar{1}10]}/dH_{[\bar{1}10]}$ increases linearly up to $H_C$ and reveals a kink at $H_C$, after which it begins to decrease. The temperature evolution of the nonlinear magnetoelectric effect is presented in the Supplementary Information.

The $T$-dependence of $\varepsilon'_{[110]}$ measured at $H_{[001]} = 9$ T also shows a very sharp peak at 20.22 K (Fig. 6(a)). The variation in the magnitude of $\varepsilon'_{[110]}$ at the peak maximum is found to be 5.9%, which is approximately twice as the value found for $H_{[\bar{1}10]} = 9$ T. However, the FWHM is 0.08 K, which is the same as that for $H_{[\bar{1}10]} = 9$ T. When the $\varepsilon'_{[110]}$ is plotted in the narrow $T$ regime for $H_{[001]} = 1\text{-}9$ T (Fig. 6(b)), the peak of $\varepsilon'_{[110]}$ gradually moves to slightly higher $T$ with continuous suppression of the peak height. In contrast to the behavior of $\varepsilon'_{[110]}$ at $H_{[\bar{1}10]}$, a small peak remains at $H_{[001]} = 1$ T, with the shift of peak height from 9 T to 1 T only approximately 0.2 K. Figure 7 shows the $H_{[001]}$ dependences of $P_{[110]}$, $MD_{[110]}$, and $dM_{[001]}/dH_{[001]}$. The $P_{[110]}$ at $H_{[001]} = 9$ T is found to be 79.1 μC/m² which is approximately 1.5 times the value of $P_{[110]}$ at $H_{[\bar{1}10]} = 9$ T (Fig. 5(a)). As the $H_{[001]}$ is decreased from 9 T, the $P_{[110]}$ decreases monotonously with a slight change of its slope between 6 and 3 T. Further decrease in $H_{[001]}$ induces continuous reduction of the $P_{[110]}$, resulting in an antisymmetric field dependence of $P_{[110]}$ for negative values of $H_{[001]}$. The $MD_{[110]}$, shown in Fig. 7(b) increases slightly from 9 T and a small change in slope occurs below 6 T. It decreases monotonously below 3 T, accompanied by symmetric behavior for negative $H_{[001]}$ values, with approximately 0.4% variation in the whole measurement range of $H_{[001]}$. The $dM_{[001]}/dH_{[001]}$ varies smoothly in the entire $H_{[001]}$ range, dissimilar from the distinct anomaly observed at the spin-flop transition for the $H_{[\bar{1}10]}$ direction.

Despite the presence of several off-diagonal components in the magnetoelectric tensor of CTO, the dominant $H$-driven change of polarization occurs for the $P_{[110]}$, consistent with the results of recent theoretical calculations based on the effective Hubbard model[6]. The trigonal $P\bar{3}c1$ structure in CTO involves two types of inequivalent sublattices for the magnetic $Co^{2+}$ ions. The theoretical work presents an important consequence that each magnetic sublattice produces electric polarization with a different magnitude and direction, each of which varies linearly with the applied magnetic field strength. Thus, the superposition of two different contributions leads to the observed nonlinear behavior in the total polarization. The largest value of $P_{[110]}$ at $H_{[001]} = 9$ T and its almost linear variation upon increasing $H_{[001]}$ seems to be compatible with the theoretical estimations. However, the highly-nonlinear magnetoelectric effect in the $P_{[110]}$ under $H_{[\bar{1}10]}$ and $H_{[110]}$ implies significant contribution of each sublattice to the magnetic-field dependent polarization. The absence of the $P_{[110]}$ within the low $H_{[\bar{1}10]}$ or $H_{[110]}$ regime (below the spin-flop transition) suggests that two different contributions compensate each other completely. Above the transition, the dominant negative-polarization arising from one sublattice gives rise to the negative net $P_{[110]}$, but the gradual increase of the positive-polarization from the other sublattice results in the broad minimum and further increase of the net $P_{[110]}$ upon increasing the field. Our results motivate more elaborate theoretical calculations comprising other factors such as additional lattice contributions and possible change of magnetic structure driven by electric field poling, which have not been considered in the previous studies.

**Conclusion**

In summary, we have synthesized single crystals of magnetoelectric $Co_4Ta_2O_9$ and explored magnetic and magnetoelectric properties along different crystallographic orientations. An in-plane antiferromagnetic order of $Co^{2+}$ moments sets in at $T_N$ = 20.5 K, below which ferroelectricity appears principally along the [110] axis under applied magnetic fields. Different from the linear magnetoelectricity in other isostructural compounds, antisymmetric and nonlinear magnetoelectric effects are demonstrated for magnetic fields applied along in-plane directions. The polarization along the [110] axis is absent at zero in-plane magnetic field, and starts to decrease to negative values above the spin-flop transition at $H_C \approx 0.3$ T. Further increase of the magnetic field leads to a broad (and negative) minimum, followed by a gradual increase of the polarization to approximately ~60 μC/m² at 9 T. Our discovery of the nonlinear magnetoelectric behavior above the spin-flop transition indicates the complex evolution of polarization components with opposite signs from different $Co^{2+}$ sublattices. These results provide vital insights into fundamental magnetoelectric interactions, paving way for the discovery of novel materials for magnetoelectric functional applications.

**Methods**

Hexagonal plate-like single crystals of CTO were grown by the conventional flux method with NaF, $Na_2CO_3$, and $V_2O_5$ fluxes in air[32]. $Co_3O_4$, and $Ta_2O_5$ powders were mixed in the stoichiometric ratio and ground in a mortar, followed by pelletizing and calcining at 900 °C for 10 h in a box furnace. The calcined pellet was finely reground and sintered at 1000 °C for 15 h. After regrinding, the same sintering procedure was carried out at 1100 °C for 24 h. A mixture of pre-sintered polycrystalline powder and fluxes was heated to 1280 °C in a Pt crucible. It was melted at the soaking $T$, slowly cooled to 800 °C at a rate of 1 °C/h, and cooled to room $T$ at a rate of 100 °C/h.

The $T$ and $H$ dependences of the DC $M$ were measured using a vibrating sample magnetometer at $T$ = 2-300 K and $H$ = –9-9 T in a physical properties measurement system (PPMS, Quantum Design, Inc.). The $C$ was measured with the standard relaxation method in the PPMS. The $T$ dependence of the AC $M$ with various frequencies was obtained in a magnetic properties measurement system (Quantum Design, Inc.). The $T$ and $H$ dependences of $\varepsilon'$ were observed at $f$ = 100 kHz using an LCR meter (E4980, Agilent). The $T$ and $H$ dependences of electric polarization ($P$) was obtained by integrating pyro- and magneto-electric currents, respectively, measured after poling in a static electric field.

**Acknowledgements**

The work at Yonsei was supported by the National Research Foundation of Korea (NRF) Grants (NRF-2017K2A9A2A08000278, NRF-2017R1A5A1014862 (SRC program: vdWMRC center), NRF-2018R1C1B6006859, and NRF-2019R1A2C2002601). HYC was partially supported by the Graduate School of YONSEI University Research Scholarship Grants in 2017. VK and SC were supported by the NSF, Grant No DMR-1609935. Access to the X-ray facilities in the Research Complex at Harwell is gratefully acknowledged.

**Additional information**

**Supplementary information** accompanies this paper.
**Competing interests:** The authors declare no competing interests.

**Figures and Captions**

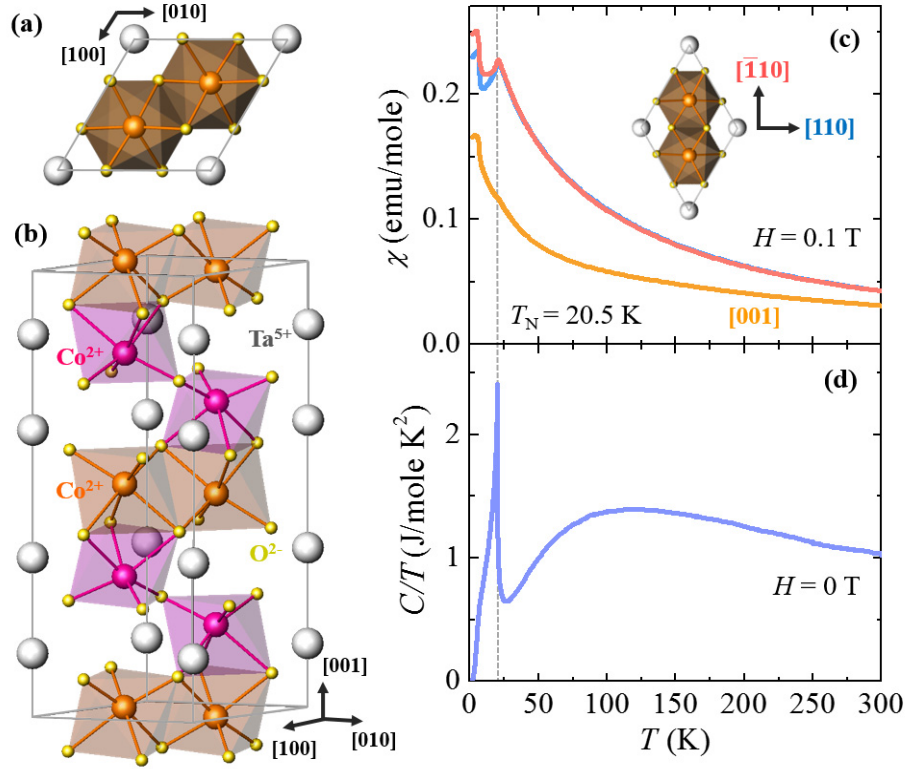

**Figure 1. Crystallographic structure and temperature dependence of magnetic properties.** (a) and (b) Views of the crystal structure of $Co_4Ta_2O_9$ ($P\bar{3}c1$) from the top (a) and from the side (b). Orange and pink spheres represent two inequivalent $Co^{2+}$ ions, and light grey and yellow spheres denote $Ta^{5+}$ and $O^{2-}$ ions, respectively. The grey box with the rhombic cross-section represents the crystallographic unit cell. (c) Temperature dependence of magnetic susceptibility, $\chi=M/H$, at $H$ = 0.1 T applied along the three inequivalent crystallographic orientations [$\bar{1}10$], [110] and [001]. Inset shows the in-plane directions for the $\chi$ measurement. (d) Temperature dependence of specific heat divided by the temperature, $C/T$, measured at the zero magnetic field. Dotted line indicates the Néel temperature $T_N$ = 20.5 K.

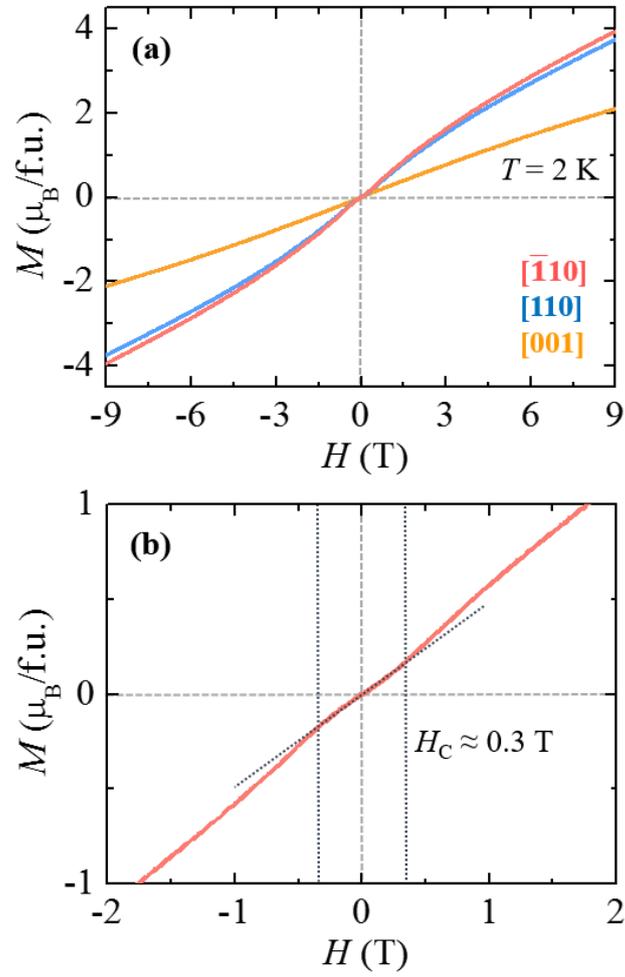

**Figure 2. Anisotropic isothermal magnetization for Co$_4$Ta$_2$O$_9$.** (a) Isothermal magnetization, $M$, measured at 2 K in the magnetic field range of ±9 T along the [$\bar{1}$10], [110] and [001] axes. (b) Magnified plot of $M$ for the [$\bar{1}$10] direction. Short-dotted vertical lines indicate metamagnetic transitions occurring at $H_C \approx \pm 0.3$ T

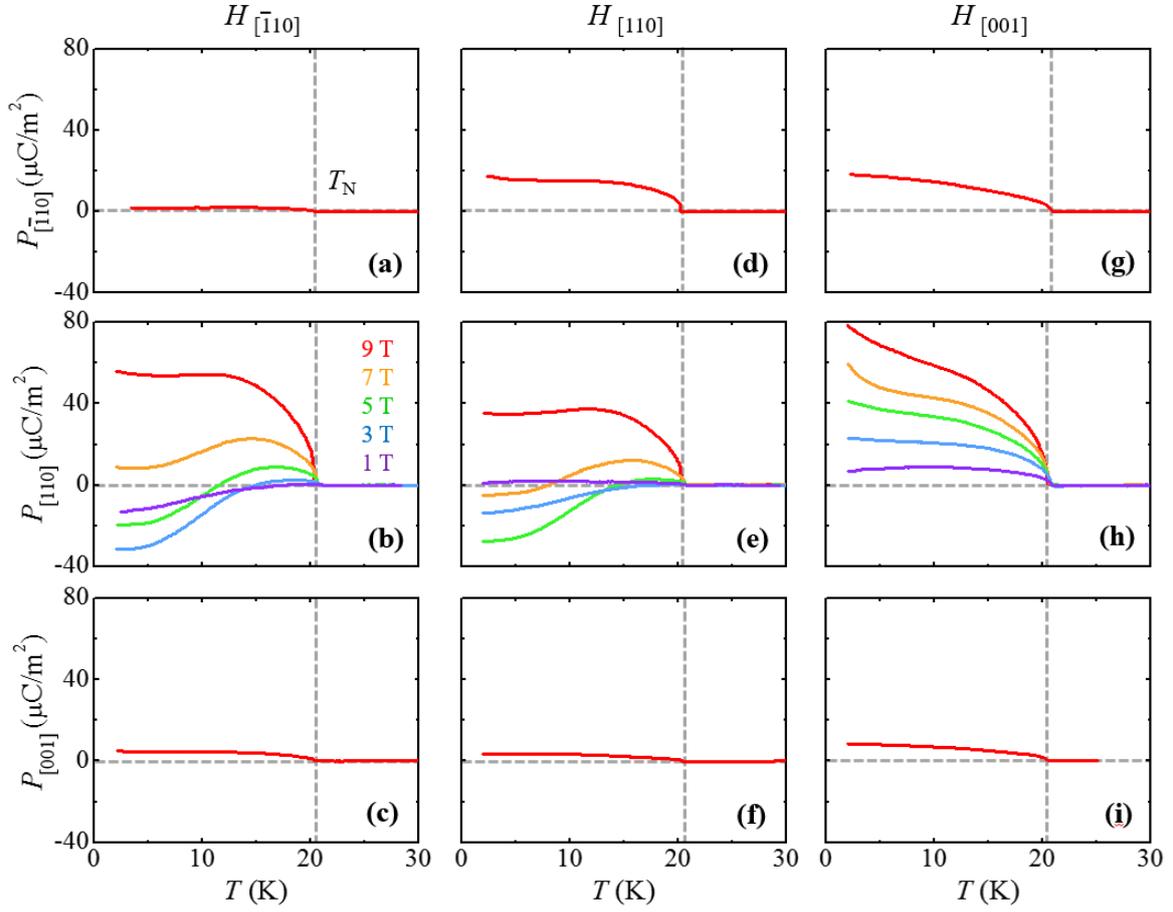

**Figure 3. Temperature dependence of the anisotropic ferroelectric polarization.** (a)-(c) Temperature dependence of $P_{[\bar{1}10]}$, $P_{[110]}$, and $P_{[001]}$, respectively, obtained by integrating the pyroelectric current after poling from 100 K to 2 K in $H$. $P_{[110]}$ was measured at $H_{[\bar{1}10]}$ = 1, 3, 5, 7 and 9 T, and $P_{[\bar{1}10]}$ and $P_{[001]}$ were measured at $H_{[\bar{1}10]}$ = 9 T. (d)-(f) Temperature dependence of $P_{[\bar{1}10]}$, $P_{[110]}$, and $P_{[001]}$, respectively, at $H_{[110]}$. (g)-(i) Temperature dependence of $P_{[\bar{1}10]}$, $P_{[110]}$, and $P_{[001]}$, respectively, at $H_{[001]}$.

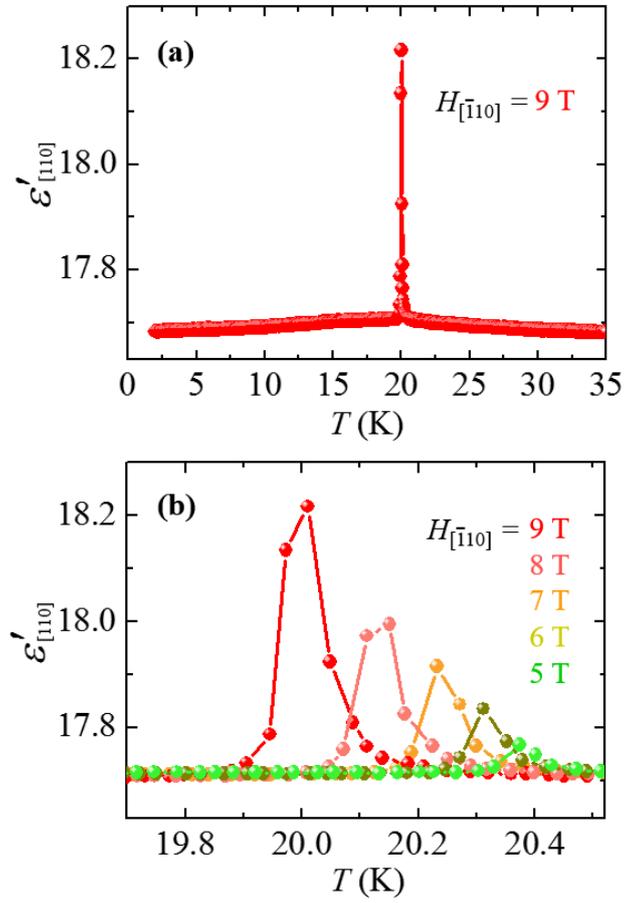

**Figure 4. Dielectric constant along the [110] axis at $H_{[\bar{1}10]}$.** Temperature dependence of dielectric constant, $\varepsilon'_{[110]}$, below 35 K at $H_{[\bar{1}10]}$ = 9 T. (b) Temperature dependence of $\varepsilon'_{[110]}$ in a narrow range of $T$ near $T_N$ at $H_{[\bar{1}10]}$ = 5, 6, 7, 8, and 9 T.

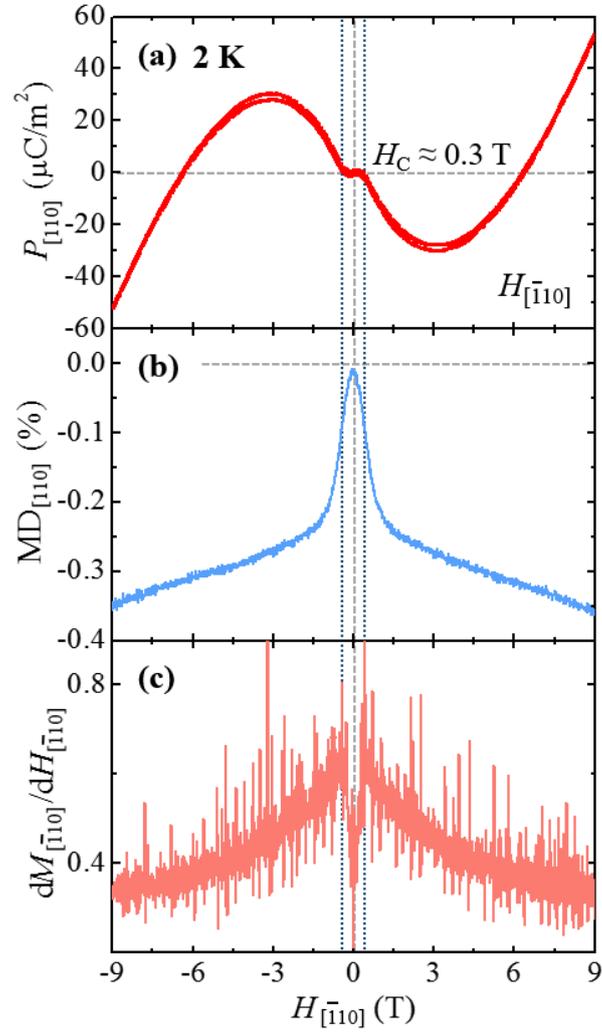

**Figure 5. Comparison of electric and magnetic properties.** (a) $H_{[\bar{1}10]}$ dependence of $P_{[110]}$ at $T = 2$ K. (b) $H_{[\bar{1}10]}$ dependence of the magnetodielectric effect along the [110] axis, $MD_{[110]}$ (%) = $\frac{\varepsilon'(H) - \varepsilon'(0\ \text{T})}{\varepsilon'(0\ \text{T})} \times 100$, measured with AC excitation of $E_{[110]} = 1$ V at $f = 100$ kHz and $T = 2$ K. (c) $H_{[\bar{1}10]}$ derivative of $M_{[\bar{1}10]}$ at 2 K.

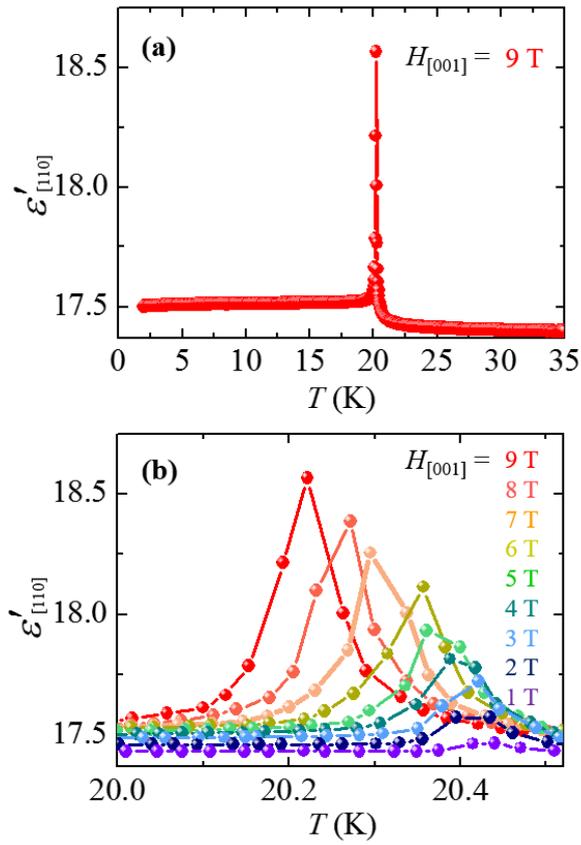

**Figure 6. Dielectric constant along the [110] axis at $H_{[001]}$.** Temperature dependence of dielectric constant, $\varepsilon'_{[110]}$, below 35 K at $H_{[001]}$ = 9 T. (b) Temperature dependence of $\varepsilon'_{[110]}$ in the narrow range of $T$ near $T_N$ at $H_{[001]}$ = 1-9 T.

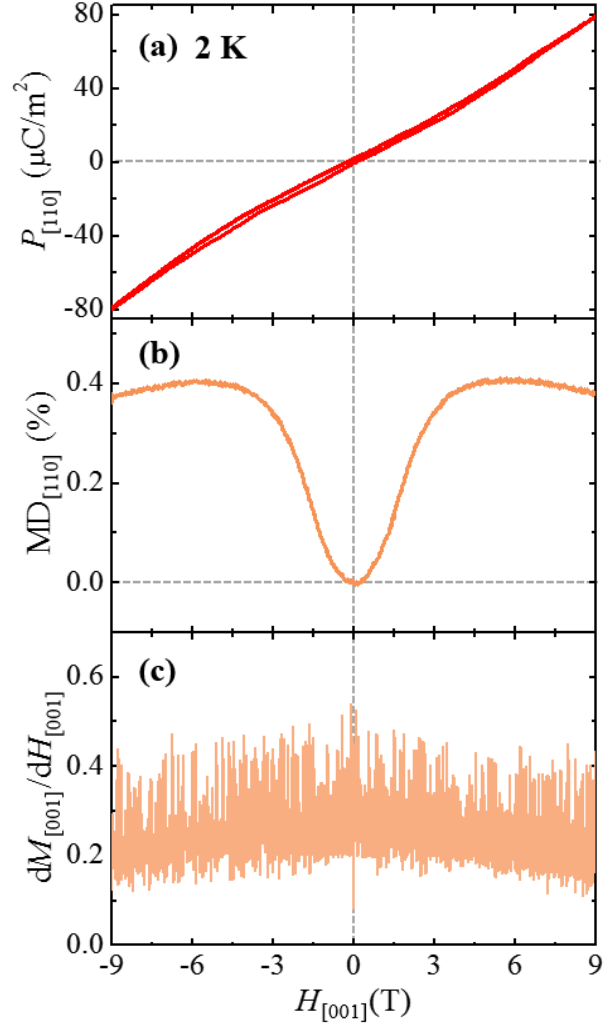

**Figure 7. Comparison between electric and magnetic properties.** (a) $H_{[001]}$ dependence of $P_{[110]}$ at $T = 2$ K. (b) $H_{[001]}$ dependence of the magnetodielectric effect along the [110] axis, $\mathrm{MD}_{[110]}$ (%) $= \frac{\varepsilon'(H)-\varepsilon'(0\ \mathrm{T})}{\varepsilon'(0\ \mathrm{T})} \times 100$, measured with AC excitation of $E_{[110]} = 1$ V at $f = 100$ kHz and $T = 2$ K. (c) $H_{[001]}$ derivative of $M_{[001]}$ at 2 K.

# Supplementary Information for **Highly nonlinear magnetoelectric effect in antiferromagnetic $Co_4Ta_2O_9$ single crystals**

Nara Lee,[1] Dong Gun Oh,[1] Sungkyun Choi,[2,3] Jae Young Moon,[1] Jong Hyuk Kim,[1] Hyun Jun Shin,[1] Hwan Young Choi,[1] Kwanghyo Son,[4] Matthias J. Gutmann,[5] Gideok Kim,[2] Jürgen Nuss,[2] Valery Kiryukhin[3] and Young Jai Choi[1,*]

[1]*Department of Physics, Yonsei University, Seoul 03722, Korea*
[2]*Max Planck Institute for Solid State Research, Heisenbergstrasse 1, 70569 Stuttgart, Germany*
[3]*Department of Physics and Astronomy, Rutgers University, Piscataway, New Jersey 08854, USA*
[4]*Department of Modern Magnetic Systems, Max Planck Institute for Intelligent Systems, Heisenbergstrasse 3, D-70569 Stuttgart, Germany*
[5]*Rutherford Appleton Laboratory, ISIS Facility, Chilton Didcot, Oxfordshire OX11 0QX, United Kingdom*

These authors contributed equally to the work: Nara Lee, Dong Gun Oh
[*] Correspondence and request for materials should be addressed to Y. J. C. (phylove@yonsei.ac.kr)

## S1. Single-crystal X-ray diffraction

Crystal, which is suitable for single-crystal X-ray diffraction, was selected under high viscosity oil, and mounted with some grease on a loop made of Kapton foil (Micromounts™, MiTeGen, Ithaca, NY). Diffraction data were collected at 298 K with a SMART APEXII CCD X-ray diffractometer (Bruker AXS, Karlsruhe, Germany), using graphite-monochromated MoK$\alpha$ radiation. The reflection intensities were integrated with the SAINT subprogram in the Bruker Suite software[1], a multi-scan absorption correction was applied, using SADABS[2]. The structure was solved by direct methods and refined by full-matrix least-square fitting with the SHELXTL software package[3,4]. Crystal data and data collection details are given in Table S1, atomic coordinates in Table S2. Further details of the crystal structure investigation may be obtained from the Fachinformationzentrum Karlsruhe, D-76344 Eggenstein-Leopoldshafen, Germany, on quoting the depository number (http://www.fiz-karlsruhe.de/), see Table S1.

## S2. AC magnetic susceptibility of $Co_4Ta_2O_9$ at $H_{[\bar{1}10]}$

Figure S1(a) shows the temperature ($T$) dependence of the DC magnetic susceptibility ($\chi$), measured at $H = 0.1$ T along the $[\bar{1}10]$ axis ($H_{[\bar{1}10]}$), on warming after zero-field-cooling (ZFC) and cooling in the same field (FC). A sudden increase of $\chi$ occurs with the maximum slope determined by the $T$ derivative of $\chi$ at $T_C = 6.5$ K. Despite the distinct anomaly of $\chi$ at $T_C$, only a slope change is found in the heat capacity ($C$) divided by the $T$, $C/T$, as shown in Fig. S1(b). This implies the formation of a short-range ferromagnetic phase accompanied by the possible generation of a glassy state[5,6]. The $T$ at which the ZFC and FC $\chi$ curves begin to split was observed at $T_f = 5.5$ K, indicative of the onset of magnetic irreversibility. The detailed characteristics of the transition were investigated by the $T$ dependence of real ($\chi'$) and imaginary ($\chi''$) parts of AC $\chi$, measured using an AC excitation field $H_{[\bar{1}10]} = 10$ Oe with various frequencies, $f = 1, 50, 250$ and $1000$ Hz. The measured $\chi'$ and $\chi''$ results are shown in Figs. S1(c) and (d), respectively. At $T_N$, both $\chi'$ and $\chi''$ do not depend on $f$. In contrast to this behavior, the $\chi'$ near $T_f$ shows a definite $f$ dependence as the peak shifts to a higher $T$ with a lower magnitude with increasing $f$. Moreover, there is a discernible dissipative behavior seen in $\chi''$. The irreversibility of $\chi$ between ZFC and FC, and the $f$ dependence of $\chi'$ strongly indicates the onset of an additional glassy state accompanied by a small ferromagnetic component below $T_f$.

## S3. Temperature evolution of polarization ($P_{[110]}$) and isothermal magnetization ($M_{[\bar{1}10]}$) in $H_{[\bar{1}10]}$

The $T$ evolution of strongly nonlinear magnetoelectric effect in $Co_4Ta_2O_9$ was examined by comparison among isothermal $P_{[110]}$, $M_{[\bar{1}10]}$, and $dM_{[\bar{1}10]}/dH_{[\bar{1}10]}$ at $H_{[\bar{1}10]}$ up to $\pm 9$ T and $T = 5$, 10, 15, and 20 K below $T_N$, as plotted in Fig. S2. At 5 K, the overall $H_{[\bar{1}10]}$ dependences of $P_{[110]}$ and $M_{[\bar{1}10]}$ tend to behave akin to those at 2 K (Fig. 5 of the main manuscript). In comparison with the $P_{[110]}$ at 2 K (Fig. 5(a) of main manuscript), the maximum value of $P_{[110]}$ at 5 K and 9 T reduces slightly to 45.1 μC/m$^2$ (Fig. S2(a)) and the $M_{[\bar{1}10]}$ at 9 T also decreases to ~3.72 μ$_B$/f.u. (Fig. S2(e)). Upon decreasing $H_{[\bar{1}10]}$, a broad minimum of the $P_{[110]}$ (= −31.8 μC/m$^2$) occurs at 3.1 T (Fig. S2(a)) and the d$M_{[\bar{1}10]}$/d$H_{[\bar{1}10]}$ at 5 K reveals kinks at $H_C$ = ±0.3 T (Fig. S2(i)), consistent with the plateau region within $H_C$ in the $P_{[110]}$ curve. At 10 K, the broad minimum of $P_{[110]}$ occurs at 2.9 T with a significantly reduced value of −8.1 μC/m$^2$ (Fig. S2(b)). However, the maximum value of $P_{[110]}$ = 58.9 μC/m$^2$ at 9 T is found to be the largest despite the slight decrease of $M_{[\bar{1}10]}$ (~3.64 μ$_B$/f.u., Fig. S2(f)). At 15 K, the regime of nearly zero $P_{[110]}$ extends up to ±3.0 T with the absence of the broad minimum (Fig. S2(c)). At 20 K, the $P_{[110]}$ almost disappears (Fig. S2(d)) throughout the measurement region of $H_{[\bar{1}10]}$ while the $M_{[\bar{1}10]}$ shows a linear increase upon increasing $H_{[\bar{1}10]}$ and finally becomes ~3.50 μ$_B$/f.u. at 9 T (Fig. S2(h)).

**Table S1.** Crystallographic data of $Co_4Ta_2O_9$, as obtained from single-crystal X-ray diffraction at 298 K.

|  | $Co_4Ta_2O_9$ |
|---|---|
| **Formula weight** | 741.62 |
| **Crystal shape, color** | block, black |
| **Space group (no.), Z** | $P\bar{3}c1$ (no. 165), 2 |
| **Lattice parameters /Å** | $a$ = 5.1718(13) |
|  | $c$ = 14.127(3) |
|  | $c/a$ = 2.732 |
| **V /Å³** | 327.3(2) |
| **$\rho_{xray}$ /g×cm⁻³** | 7.526 |
| **Crystal size /mm³** | 0.15×0.10×0.05 |
| **Diffractometer** | SMART APEX II, Bruker AXS |
| **X-ray radiation, $\lambda$/Å** | MoK$\alpha$, 0.71073 |
| **Absorption correction** | Multi-scan, SADABS[2] |
| **2$\theta$ range /°** | 5.76° ≤ 2$\theta$ ≤ 72.84° |
| **Index ranges** | -8 ≤ $h$ ≤ 8 |
|  | -8 ≤ $k$ ≤ 8 |
|  | -22 ≤ $l$ ≤ 23 |
| **Reflections collected** | 5468 |
| **Data, $R_{int}$** | 540, 0.0409 |
| **No. of parameters** | 25 |
| **Transmission: $t_{min}$, $t_{max}$** | 0.034, 0.110 |
| **Extinction coefficient** | 0.0241(7) |
| **Final R indices [I > 2$\sigma$(I)]** | R1 = 0.0158, wR2 = 0.0364 |
| **R indices (all data)** | R1 = 0.0165, wR2 = 0.0378 |
| **Deposition no.** | CSD-?????? |

**Table S2.** Atomic coordinates, equivalent isotropic displacement factors (Å$^2$) for Co$_4$Ta$_2$O$_9$ at 298 K.

| atom | site | $x$ | $y$ | $z$ | $U_{eq}$ |
|---|---|---|---|---|---|
| Ta | 4$c$ | 0 | 0 | 0.14280(2) | 0.00843(7) |
| Co1 | 4$d$ | 1/3 | 2/3 | 0.19186(4) | 0.0103(1) |
| Co2 | 4$d$ | 1/3 | 2/3 | −0.01396(4) | 0.0100(1) |
| O1 | 6$f$ | 0 | 0.2870(3) | 1/4 | 0.0102(3) |
| O2 | 12$g$ | 0.0230(3) | 0.6814(3) | 0.0844(1) | 0.0119(2) |

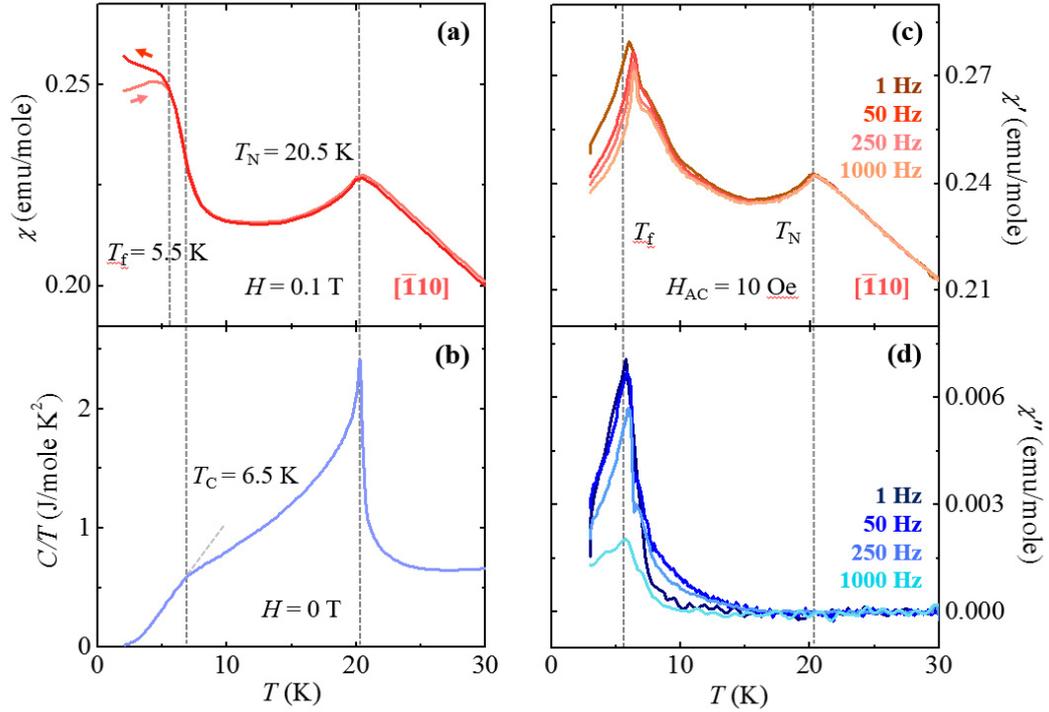

**Figure S1. Comparison between DC and AC magnetic susceptibilities.** (a) Temperature ($T$) dependence of the magnetic susceptibility, $\chi = M/H$, of Co$_4$Ta$_2$O$_9$ at $H_{[\bar{1}10]} = 0.1$ T, measured upon warming after zero-field cooling and upon cooling. The vertical dashed lines indicate the antiferromagnetic transition, short-range ferromagnetic transition, and freezing temperatures, $T_N = 20.5$ K, $T_C = 6.5$ K and $T_f = 5.5$ K, respectively (b) $T$ dependence of heat capacity divided by the temperature, $C/T$, measured in zero $H$ up to 30 K. (c)-(d) $T$ dependence of real ($\chi'$) and imaginary ($\chi''$) parts of AC $\chi$, measured at various frequencies, $f = $ 1, 50, 250 and 1000 Hz, under zero DC $H$ and AC field excitation of 10 Oe ($H_{AC} = 10$ Oe) along the $[\bar{1}10]$ axis.

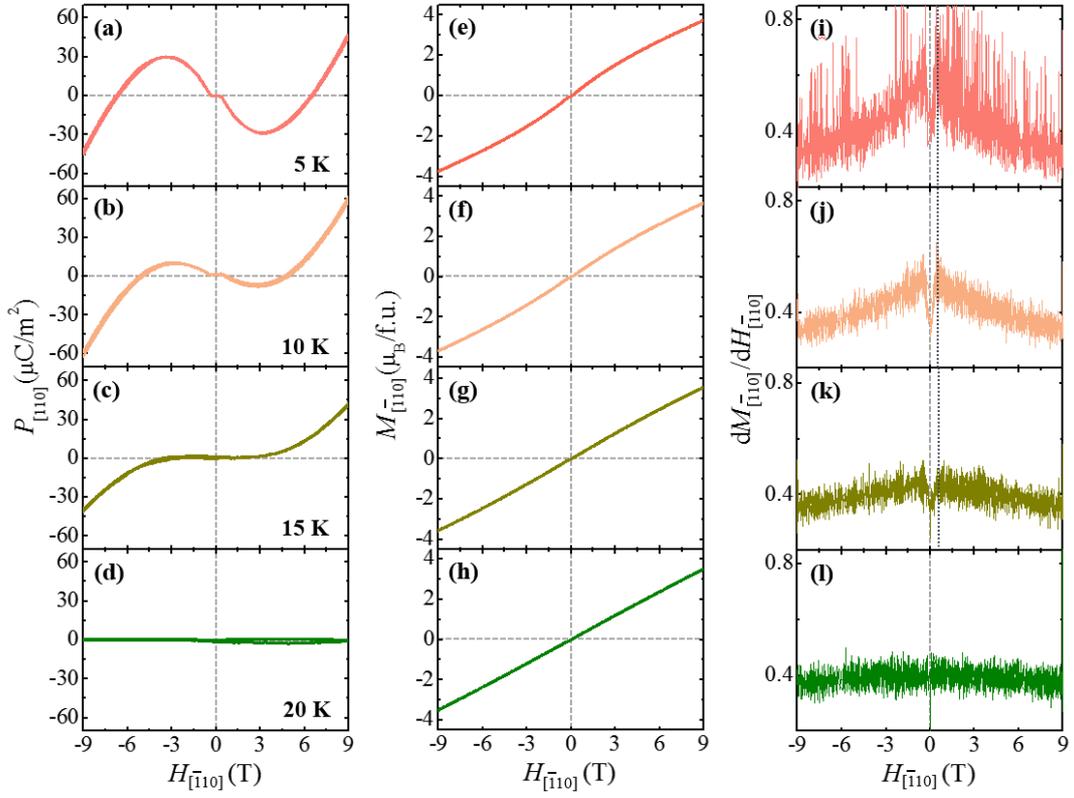

**Figure S2. Temperature evolution of ferroelectric polarization and magnetization.** (a)-(d) $H_{[\bar{1}10]}$ dependence of ferroelectric polarization ($P_{[110]}$) at $T$ = 5, 10, 15 and 20 K, respectively, obtained by integrating the magnetoelectric current measured changing $H_{[\bar{1}10]}$ at the rate of 0.01 T/s up to ±9 T after poling in $E_{[110]}$ = 4.72 kV/cm and $H_{[\bar{1}10]}$ = 9 T. (e)-(h) $H_{[\bar{1}10]}$ dependence of magnetization ($M_{[\bar{1}10]}$) at $T$ = 5, 10, 15 and 20 K, respectively, measured up to ±9 T. (i)-(l) $H_{[\bar{1}10]}$ derivative of $M_{[\bar{1}10]}$ at $T$ = 5, 10, 15 and 20 K, respectively.